\documentclass[twocolumn,showpacs,preprintnumbers,amsmath,amssymb,floatfix]{revtex4}

\usepackage{epsfig}% Include eps figure files
\usepackage{dcolumn}% Align table columns on decimal point
\usepackage{bm}% bold math

\begin{document} 

\title{Critical Survey of
Isoscalar and Isovector Contributions to the Spin Orbit Potential
in Relativistic Mean Field Theory}

\author{A. Bhagwat$^{1}$, R. Wyss$^{1}$, W. Satu{\l}a$^{1,2}$, 
       J. Meng$^{3}$ and Y. K. Gambhir$^{4}$ }

\affiliation{$^{1}$ Royal Institute of Technology (KTH),
Department of Nuclear Physics, AlbaNova University Center, S-106 91 Stockholm,
Sweden \\
$^{2}$ Institute of Theoretical Physics, University of Warsaw,
ul. Hoza 69, PL-00 681 Warsaw, Poland \\
$^{3}$ School of Physics, Peking University, Beijing 100871, China \\
$^{4}$ Department of Physics, I.I.T. Powai, Bombay 400076, India}

\date{\today}

\begin{abstract}
The spin-orbit (SO) interaction, emerging naturally from the Relativistic Mean
Field (RMF) theory is examined critically in the light of the recently measured
excitation energy differences between the terminating states built on two
different configurations for nuclei belonging to the lower pf shell.
The calculations are carried out using the cranked RMF framework. To further
probe the iso-vector dependence of the spin-orbit potential, the energy
spacing between the $g_{7/2}$ and  $h_{11/2}$ states in the Sb-chain is compared
to experiment. It is found that the calculation at the quantitative level
deviates strongly from the experiment. In particular the balance of the 
iso-scalar and iso-vector strengths of the effective one body SO potential 
indicates that additional terms like tensor couplings may be needed to account 
for the experimental data.
\end{abstract}

%\pacs{21.10.-k,21.30.Fe,21.60.-n}

\maketitle

Low energy excitations and global properties in atomic nuclei are rather well
described by the modern effective forces.
Different realizations of the short range
nuclear interaction have been developed, like the Skyrme and Gogny for non
relativistic Hartree Fock potentials and the Relativistic Mean Field (RMF)
Lagrangian, based on meson couplings. These effective theories employ a set
of parameters that are adjusted to selected experimental observables in order
to account for basic nuclear structure properties. Since there exist a
multitude of observables and no real consensus about which of these are to be
selected in a unique manner, there exist a multitude of different force
parameterizations. One of the basic problems in adjusting the force parameters
is related to the fact that the single particle states, that are so
crucial for high accuracy calculations, are in general coupled to collective
motion and therefore difficult to determine. In this context, any
dataset representing single-particle motion
manifests itself as an invaluable source
for a rigorous test and subsequent fine-tuning of the parameters.

\begin{table*}[htb]
\caption{Summary of experimental results on terminating states in
lower pf shell.}
\label{tab1}
\newcommand{\m}{\hphantom{$-$}}
\newcommand{\cc}[1]{\multicolumn{3}{|c|}{#1}}
\renewcommand{\arraystretch}{1.2}
\begin{tabular}{c|rcc|rcc|c}
\hline
          & \cc{$f^{n}_{7/2}$}  &\cc{$d^{-1}_{3/2}f^{n+1}_{7/2}$} \\ \cline{2-4}\cline{5-7}
Nucleus   &$I_{max}$ &$E_{max}$ (MeV)& Ref.    &$I_{max}$ &$E_{max}$ (MeV)&Ref.         &$\Delta E$ (MeV)\\
%          &          & (MeV)   &         &          & (MeV)   &    & (MeV) \\
[1pt]\hline
%                   &       &             &        &       &             &    \\
$^{42}$Ca &6$^+$   & 3.189 &\cite{LAC.03}&11$^-$  & 8.297 &\cite{LAC.03}& 5.108 \\
$^{44}$Ca &8$^+$   & 5.088 &\cite{LAC.01}&13$^-$  &10.568 &\cite{LAC.01}& 5.480 \\
$^{44}$Sc &11$^+$  & 3.567 &\cite{LAC.05}&15$^-$  & 9.141 &\cite{LAC.05}& 5.574 \\
$^{45}$Sc &23/2$^-$& 5.417 &\cite{BED.01}&31/2$^+$&11.022 &\cite{BED.01}& 5.605 \\
$^{45}$Ti &27/2$^-$& 7.144 &\cite{BED.98}&33/2$^+$&13.028 &\cite{BED.04}& 5.884 \\
$^{46}$Ti & 14$^+$ &10.040 &\cite{FB.04} &17$^-$  &15.550 &\cite{FB.04} & 5.510 \\
$^{47}$V&31/2$^-$&10.004 &\cite{FB.01} &35/2$^+$&15.259 &\cite{FB.01} & 5.255\\
%                 &       &             &        &       &             &  \\
\hline
\end{tabular}
\end{table*}

Recently, it was shown \cite{Zdu05,WS.05} that a
set of terminating states in the mass region A$\sim$ 45 may provide a unique
tool to determine the strength of the spin-orbit (SO) interaction. The long 
experience
in the studies of terminating states within the Nilsson model have clearly
shown that these belong to the purest single particle states available in
nuclear structure \cite{AFA.99,SAT.05} and hence provide an excellent
laboratory for the adjustment of effective forces. Indeed, in Refs.
\cite{Zdu05,WS.05} it was demonstrated that for some of the Skyrme
force parameterizations, the SO interaction can be adjusted in such a
way, that it agrees with the experimental data, whereas some other force
parameterizations can be ruled out as irreparable. The robustness of terminating
states in determining properties of the nuclear energy density functional
was further supported by the recent comparative study between the fully
correlated shell model and the Skyrme Hartree-Fock (SHF) \cite{wojtek},
showing essentially a one to one correspondence. 

Since the
SO potential emerges naturally within the RMF
\cite{JDW.74}, it becomes highly interesting and important to carry out a
critical survey of the SO interaction within the RMF framework, in the
light of the recently measured terminating states in the lower pf shell.

In addition, there is a long standing debate on the iso-vector strength of the
SO potential for different Skyrme parameterizations, exemplified
in the recent extension \cite{Rei95} of the Skyrme SO potential,
which needs to be confronted to experiment. Similarly, in the
RMF, there is an open question as to what extend the Hartree potential that
generates the SO field is sufficient to account for the experimental
data or whether additional terms or constraints, e.g. tensor coupling 
\cite{JIA.05} need to be added in the theoretical framework.
 
One may also consider the Fock terms involving the $\pi$ meson,
which strongly affect the effective one body SO potential \cite{Bou87}.
In this work, we carry out a systematic investigation of the
SO potential by comparison to the recent data from the lower pf
shell within the framework of the cranked RMF (c-RMF) \cite{KOE.89,AFA.96}.
To further address the iso-vector dependence of the SO potential,
we employ the recently studied $g_{7/2}-h_{11/2}$ spacing,
$E^{gh}$, in the chain of Sb isotopes \cite{SCH.04}. 

Recently, experimental data on the terminating states in the lower pf shell
has been reported in the literature
\cite{LAC.03,LAC.01,LAC.05,BED.01,BED.98,BED.04,FB.04,FB.01}. The terminating
states considered here involve the $f^{n}_{7/2}$ as well as the
$d_{3/2}^{-1}f^{n+1}_{7/2}$ configurations where $n$ denotes the number of
valence particles outside the $^{40}$Ca core and the particle hole excitation
across the magic gap 20 is of proton type. Summary of the relevant experimental
results 
is presented in Table \ref{tab1}.
Following the discussion in Refs.
\cite{Zdu05,WS.05}, we extract the difference, $\Delta E$, between
the excitation energies of the states terminating within the
$d_{3/2}^{-1}f_{7/2}^{n+1}$ and $f_{7/2}^n$ configurations. The value
of $\Delta E$ depends predominantly on the size of the magic gap 20
which in turn directly relates to the strength of the SO potential.
Within the spherical Nilsson
Hamiltonian \cite{Nil55} the magic gap
$  \Delta e_{20} = \hbar\omega_o  ( 1 - 6\kappa - 2\kappa\mu )$,
depends on three major factors: i) the bulk
properties of the potential characterized by $\hbar\omega_o$, ii) the strength
of the SO term $\kappa$ and  iii) the flat-bottom and surface
properties entering through the orbit-orbit term, $\sim\mu$. All three terms
influence the magic gap 20 with a well defined hierarchy. For light nuclei one
can disregard the flat-bottom effect ($\mu \sim 0$). Note that the smaller the
energy spacing between $1d_{3/2}$ and $1f_{7/2}$, the larger the effective
SO potential.

The width of the Nilsson potential, $\hbar\omega_o$,
is fitted to nuclear radii. Hence, it sets a
common energy scale for all microscopic models
in low energy nuclear physics. Consequently,
any adjustments of, in particular,
the RMF parameters leading effectively even to small variations
in $\hbar\omega_o$ in light nuclei is expected to endanger the good
agreement already achieved between the theory and the experiment
for radii and binding energies in heavy nuclei. Therefore, the
difference in the excitation energies provides a unique testing
ground for the SO interaction usually employed in
the mean field calculations.

These simple arguments are strongly supported by the self-consistent
SHF calculations \cite{Zdu05,WS.05}. In particular, in Ref.~\cite{Zdu05}
it is shown that the difference between the SHF and the corresponding
experimental values of $\Delta E$, averaged
over all the available data,
i.e. the quantity which measures the quality
of the parametrization is directly correlated with the
isoscalar-effective-mass scaled strength (true strength) of the
iso-scalar SO interaction.  This result not only
exemplifies the role of the SO potential but also
shows that for physical observables like $\Delta E$ (or $E^{gh}$)
the impact of non-local effects (effective mass) on the
HF single-particle spectra is, as expected, compensated by
the effective coupling constants in the energy density functional.

In the RMF theory, the point nucleons are assumed
to be interacting only via the c-number electromagnetic (e.m.) and the
$\sigma$, $\omega$ and $\rho$ meson fields. Within the standard
nonlinear ($\sigma$, $\omega$, $\rho$) interaction Lagrangian
\cite{YKG.90,PRI.96} the  SO
potentials for protons and neutrons ($U^{p}_{ls}$ and $U^{n}_{ls}$)
can be expressed in terms of iso-scalar and iso-vector
contributions as (see, for example, \cite{MMS.95}):
\begin{eqnarray}
U_{ls}^{p(n)}(\vec{r})&=&\frac{1}{rm^2 {m^\star}^2}\left[
\left(C_\sigma^2+C_\omega^2\right) \nabla_r \rho(\vec{r}) \right. \nonumber \\
&& \left.\hspace{2.2cm} \pm
C_\rho^2 \nabla_r \rho_{pn}(\vec{r})\right].
\label{so}
\end{eqnarray}
Here, $m$ ($m^\star$) is the nucleon (effective) mass; the constants $C_{i}$
($i = \sigma, \omega, \rho$) are defined as $C_{i}~=~mg_{i}/m_{i}$, $m_i$
being the meson masses; $\rho(\vec{r})$ is the total nucleon density (sum of
neutron and proton densities) and $\rho_{pn}(\vec{r})$ is the difference
between proton and neutron densities. Sign + (-) arises for protons (neutrons).
Clearly, the iso-scalar term of the SO interaction
is dominant in comparison with the iso-vector term. This can be substantiated
further by examining the ratio of iso-scalar to iso-vector strengths of the
SO potential in Equation (\ref{so}). For NL1 \cite{REIN.86} and 
NL3 \cite{LAL.97} Lagrangian
parameter sets, this ratio turns out to be 16.54 and 19.62, respectively. For
Lagrangian parameter sets with density dependent couplings, DD-ME1
\cite{NIK.02} and DD-ME2 \cite{LAL.05}, this ratio is typically around 20
or even larger. Further insight into the properties of the SO potential
can be gained by assuming that proton and neutron densities have the same
shape. In this case, Equation (\ref{so}) reduces to:
\begin{eqnarray}
U_{ls}^{p(n)}
=\frac{-(C_\sigma^2+C_\omega^2)A  }{rm^2 {m^\star}^2}\left[
1 \mp \lambda \frac{N-Z}{A}\right]g(r)
\label{4}
\end{eqnarray}
Here, $-g(r)$ is the derivative of the density, and is negative definite;
and $\lambda = C_\rho^2 /(C_\sigma^2+C_\omega^2)$. As expected, the spin
orbit potential decreases (increases) for protons (neutrons) as $N-Z$
increases. However, in contrast to non-relativistic models, the iso-vector
strength, $\lambda$, is at least an order of magnitude smaller, and is not an
independent quantity that can be adjusted further.

In this work, we solve the c-RMF equations using the basis expansion technique
\cite{YKG.90}. The upper and lower components of Dirac spinor as well as
the meson and e.m. fields are expanded in terms of three dimensional harmonic
oscillator basis. We use 8 fermionic and 10 bosonic shells. 
We have explicitly checked for a few representative cases that
the use of 10 fermionic and 10 bosonic shells produces essentially the 
same values of $\Delta E$. Since primary focus is on relative
quantities like $\Delta E$, inclusion of a larger number of shells
does not alter the conclusions drawn. Further, pairing correlations
are ignored. This is justified, since it is expected that the pairing
correlations are quenched for the states considered here \cite{AFA.99,SAT.05}.
Currents are crucial for the calculation of $\Delta E$  and are taken into
account appropriately.
For further technical details of the c-RMF model, see \cite{KOE.89,AFA.96}.
The calculation are carried out using two frequently used parameter
sets, NL1 \cite{REIN.86} and NL3 \cite{LAL.97}. In the calculation of $E^{gh}$,
we also employ the RMF model with density dependent meson - nucleon couplings
\cite{NIK.02,LAL.05,TW.99} using the Lagrangian parameter set DD-ME2 
\cite{LAL.05}. Preliminary results of the present calculation may be found
in Ref. \cite{BHA.06}.

\begin{figure}[htb]
\centerline{\epsfig{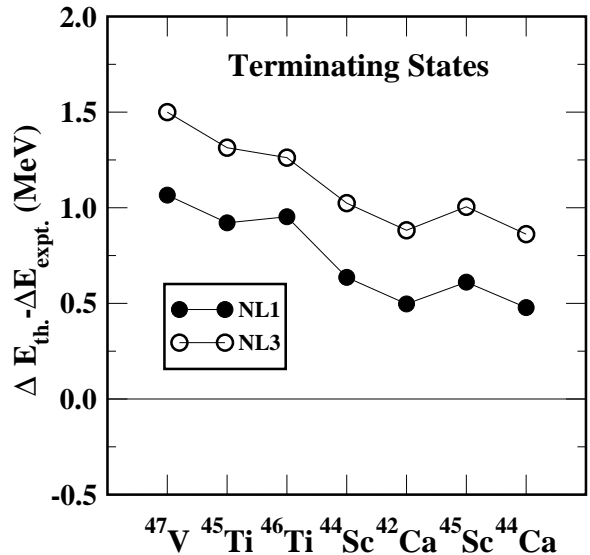}}
\caption{The difference between theoretical and the corresponding
experimental values of the excitation energy differences ($\Delta E$) for the
terminating states within the $d_{3/2}^{-1}f_{7/2}^{n+1}$ and $f_{7/2}^n$
configurations.}
\label{fig1}
\end{figure}

The difference between the calculated and the corresponding experimental
values of  $\Delta E$ is plotted in Figure \ref{fig1}. It is clear that
$\Delta E$ values obtained using the parameter sets, NL1 and NL3, though
qualitatively in agreement with the experiment, differ from each other and
from the experiment. Both over-estimate the experiment. The deviation is
in general larger than the corresponding
Skyrme calculations. Both NL1 and NL3 values
exhibit similar systematics and differ from each other roughly by 0.5 MeV
on the average. Further, the NL1 values are found to be closer to the
experiment than those obtained by using set NL3, with mean squared deviations
0.59 and 1.31 respectively. Apparently, the magnitude of the iso-scalar spin
orbit potential is at variance with experiment. It should be noted that
the nuclei in the above figure are arranged in the sequence of increasing
(N-Z)/A. Hence, a slope in  $\Delta E$ is related to inadequacy of the
iso-vector strength. An inspection of Figure \ref{fig1} reveals that for both
the parameter sets, the discrepancy between experiment and theory is maximum
for $^{47}$V and minimum for $^{44}$Ca. Thus, the discrepancy decreases with
(N-Z)/A. This analysis points towards deficiency of the SO interaction,
especially, in two ways: i) the iso-scalar term is obviously too weak ii) the
iso-vector term requires a substantial increase.

To further investigate the iso-vector part of the SO potential, we
study the chain of Sb isotopes in the RMF framework. It should be noted that
for this particular case, we solve the RMF equations without taking 
the currents into
account.
It has been shown \cite{YKG.94} that the
effect of time reversal breaking which exists in the case of odd A systems
seems to cancel while taking the differences between single particle
energies. Therefore, the conclusions drawn here will not change, even if the
currents are explicitly taken into account.

The Sb chain gives an important clue to the iso-vector dependence of
the SO potential from the spacing between single proton $2g_{7/2}$ and
$1h_{11/2}$ states, revealing that the spacing $E^{gh}$ increases
with neutron number, i.e the effective SO potential for protons
decreases with neutron number. 
%This is akin to the case of the spacing between
%the $1d_{3/2}$ and $1f_{7/2}$ states discussed earlier. 
Note that the single
particle states $2g_{7/2}$ and $1h_{11/2}$ are less pure than the terminating
states discussed above. For example, the energy spacing between $g_{7/2}$
and $h_{11/2}$ states is also influenced by
quadrupole collectivity, particularly in the mid - shell region, yielding
solutions that are somewhat deformed.
Therefore, these states do not allow an adjustment at the
same level of accuracy. Still, following the discussion in Ref.~\cite{SCH.04},
the spectroscopic factors show that these states are of dominant single
particle character, implying that they indeed can serve to probe
both iso-scalar and iso-vector strength.

\begin{figure}[htb]
\centerline{\epsfig{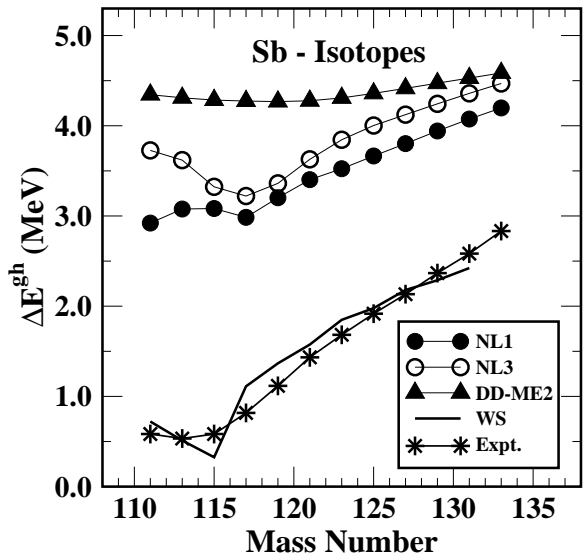}}
\caption{The calculated and the corresponding
experimental \cite{SCH.04} $E^{gh}$ values for Sb isotopes.
Solid line marks the result of microscopic-macroscopic
calculations with a Woods-Saxon potential.}
\label{fig3a}
\end{figure}

\begin{figure}[htb]
\centerline{\epsfig{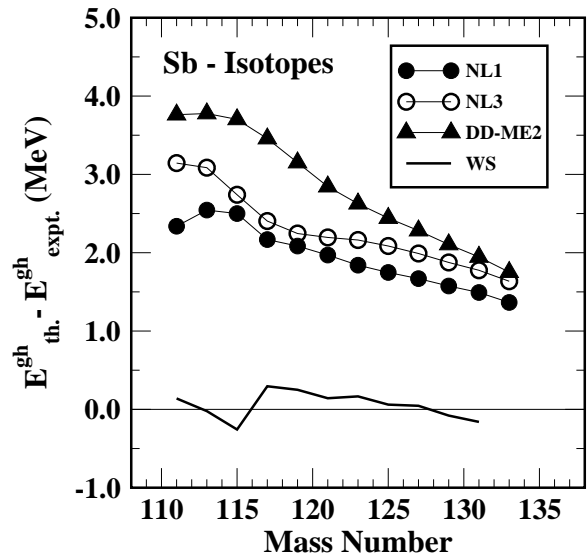}}
\caption{The difference between the calculated and the corresponding
experimental \cite{SCH.04} $E^{gh}$ values for Sb isotopes.}
\label{fig3}
\end{figure}

One should mention here that the increase of $E^{gh}$ versus $N$ is
also discussed in the literature within the context of
the nuclear shell-model, as being caused by the
monopole part of the proton-neutron interaction and is commonly dubbed
as {\it monopole migration\/} effect, see e.g. \cite{HEY.87}.%\cite{OTS.05}. 
 According to recent investigations, 
the source of {\it monopole migration\/} is a two-body
shell-model tensor interaction \cite{OTS.05}. 
The tensor
interaction can naturally be incorporated into
the energy density functional framework \cite{JIA.05,STA.77}.
%%% -- YKG:
In fact, the tensor interaction has been included in the  
tensor interaction in the recent Skyrme - Hartree - Fock studies 
\cite{BRO.06,BRI.07,COL.07}. It has been shown there that 
it plays an important role in the evolution of the shell structure,
for example, the observed trend in the $2g7/2$ - $1h11/2$ spacing.
%%% -- YKG
It is well known that in the mean-field approximation
the two-body tensor interaction leads to a one-body SO
potential, i.e. it works effectively like the
two-body SO interaction. The basic difference
between these two sources of one-body SO potential is that the
form factor generated by the two-body tensor interaction
depends on the shell-filling unlike its counterpart coming
from the two-body SO interaction. 
In an analogous manner, the $\rho NN$ tensor coupling has been
discussed for the RMF \cite{JIA.05} generating an additional contribution
to the effective one body SO potential. 
Hence, within the
mean-field approximation, the {\it monopole migration\/} can be interpreted as
a dynamical manifestation of an effective one-body SO potential. For a 
pedagogical discussion on this issue, see Ref. \cite{DOB.06}.
Consequently, with a suitable parameterization of the SO potential, one
should be able to reproduce empirical values of $E^{gh}$. In
Figures \ref{fig3a} and \ref{fig3} we demonstrate that
this is indeed the case by showing the results of a
microscopic-macroscopic calculation with a Woods-Saxon
potential, including
an iso-vector term with the same sign as in RMF but considerably
larger relative strength (see, for example, \cite{ISA.02}).

The calculated and the corresponding experimental \cite{SCH.04} values of
$E^{gh}$ are presented in Figure \ref{fig3a} for all the Lagrangian
parameter sets considered here. The figure reveals that all the calculations
deviate by large amounts from the experiment. The experimental $E^{gh}$ varies
between 0.5 MeV to around 3 MeV, whereas the corresponding calculated values
(NL1 and NL3) vary within 1 MeV only. DD-ME2, on the other hand, yields
almost a constant spacing. Note that the larger the calculated splitting,
the smaller is the effective SO potential. The difference between the
theoretical and the corresponding experimental values of  $E^{gh}$ is
presented in Figure \ref{fig3}. This figure highlights the large deviation
between theory and experiment. Interestingly, it is found that the
difference decreases with increasing mass number (hence with increasing
(N-Z)/A), for all the three parameter sets. This is similar to the case of
$\Delta E$ (see Figure \ref{fig1}).

We notice again, that the effective iso-scalar SO potential in the RMF
theory is too weak, yielding a much too large spacing between the
$2g_{7/2}$ and $1h_{11/2}$ states. Since the iso-vector dependence of the
SO is very weak in the RMF theory, with increasing (N-Z)/A, the
experimental values approach the calculated ones. By increasing the coupling
constants of the $\sigma$ and $\omega$ meson, one may increase the effective
SO potential, and at the same time keep the central potential
unchanged. However, this will only reduce the relative iso-vector
strength of the SO interaction further, see Equation (\ref{4}). 
In this way, one indeed is able to improve the effective SO potential,
however, at the same time, the iso-vector potential becomes weaker.
Our earlier study  of the symmetry energy \cite{Ban06} revealed that
the coupling strength of the $\rho$ meson is rather well adjusted to
reproduce the strength of the iso-vector central field.
Hence, our results suggest that
the presently used standard RMF Lagrangians do not allow to
simultaneously adjust the iso-vector strength, i.e. increase its value in
agreement to the observed experimental values and at the same time increase
the SO potential. 
%%% -- 
Recently, Piekarewicz \cite{PIE.07} has studied the 
single particle spectra near A = 40 by adjusting $m_\sigma$ and $g_\sigma$ 
to reproduce the  $1d_{3/2}$ - $2s_{1/2}$ energy gap in $^{40}$Ca within
the conventional RMF approach. It should be however mentioned that this is 
a purely localized study around A = 40. With only a few parameters in the 
conventional RMF,
it seems rather difficult to get a good global fit that yields the 
correct single particle spectrum, like the gap $1d_{3/2}$ - $2s_{1/2}$
or spacing $1h_{11/2}$ - $2g_{7/2}$. 
%%% --

To summarize, a systematic study of the energy splitting of the terminating
states between $f^{n}_{7/2}$ and $d_{3/2}^{-1}f^{n+1}_{7/2}$ configurations
using the cranked Relativistic Mean Field theory is carried out. The principal
aim of this work is to critically examine the SO potential, which
emerges naturally from the RMF framework. Calculations have been carried out
for NL1 and NL3 parameter sets. It is found that the one-body SO potential
is essentially too weak and at the same time
the experimentally observed iso-vector dependence stronger than what
is found in theory. The analysis of the energy spacing between the
$2g_{7/2}$ and $1h_{11/2}$ states in Sb isotopes supports these findings.
The iso-scalar part of the SO
interaction is so dominant, that there is very limited freedom to improve the
iso-vector SO potential generated by the $\rho$ meson. Even the
inclusion of density dependence of the meson - nucleon coupling strengths
does not seem to cure this problem. 
Thus, it can be concluded that the standard version
of the Lagrangian used, does not give enough freedom to achieve the desired
strength of the SO potential and the
correct iso-vector dependence in the SO interaction. This may point
to the need to extend the RMF model, for example, by including tensor 
couplings
\cite{JIA.05}
or to extend the scheme towards Hartree-Fock, including $\pi$-mesons
\cite{Bou87}.

This work has been supported by the Swedish Science Research Council (VR), the
Swedish Institute and the Polish Committee for Scientific Research (KBN)
under contract 1~P03B~059~27. 
A. B. acknowledges financial support from the Swedish Institute (SI).
The authors are thankful to P. Ring for his
interest in this work.

\end{document}